\documentclass{IEEEtran}
\usepackage{amssymb}
\usepackage{amsfonts}
\usepackage{amsmath}
\usepackage{algorithm}
\usepackage{graphicx,subfigure,cite,multicol,multirow,diagbox,booktabs,array}

\usepackage[dvips]{color}
\usepackage{float}
\ifCLASSINFOpdf
\else
\fi
\begin{document}
\title{Alternating Iterative Secure Structure between Beamforming and Power Allocation for UAV-aided  Directional Modulation Networks}

%\author{
%}% <-this % stops a space

\author{Feng Shu,~Zaoyu~Lu,~Jinyong~Lin,~Linlin Sun,~Xiaobo~Zhou,Tingting Liu,  \\Shuo Zhang, Wenlong Cai,~Jinhui Lu,~and~Jin Wang

\thanks{This work was supported in part by the National Natural Science Foundation of China (Nos. 61771244, 61501238, 61702258, 61472190, and 61271230), in part by the Open Research Fund of National Key Laboratory of Electromagnetic Environment, China Research Institute of Radiowave Propagation (No. 201500013), in part by the Jiangsu Provincial Science Foundation under Project BK20150786, in part by the Specially Appointed Professor Program in Jiangsu Province, 2015, in part by the Fundamental Research Funds for the Central Universities under Grant 30916011205, and in part by the open research fund of National Mobile Communications Research Laboratory, Southeast University, China (Nos. 2017D04 and 2013D02).}
\thanks{Feng Shu,~Zaoyu Lu,~LinLin~Sun,~Xiaobo~Zhou, Tingting Liui,~Jinhui~Lu,~and~Jin Wang are with the School of Electronic and Optical Engineering, Nanjing University of Science and Technology, 210094, CHINA. (Email: shufeng0101 @163.com). }
\thanks{Feng Shu,~and~Jin Wang are also with the School of Computer and  information at Fujian Agriculture and Forestry University, Fuzhou, 350002,  China.}
\thanks{Jinyong Lin, Shuo Zhang, and Wenlong Cai are with Beijing Aerospace Automatic Control Institute, Beijing 100854, China (e-mail: ljiny3771@sina.com)}
%\thanks{Jiangzhou Wang is with the School of Engineering and Digital Arts, University of Kent, Canterbury CT2 7NT, U.K. Email: \{j.z.wang\}@kent.ac.uk.}

}

\maketitle

\begin{abstract}
In unmanned aerial vehicle (UAV) networks, directional modulation (DM) is adopted to improve the secrecy rate (SR) performance. Alice, a ground base station, behaves as a control center, and Bob is a UAV of flying along a linear flight trajectory who optimizes its SR performance by dynamically adjusting its beamforming vectors and power allocation (PA) strategy. Per fixed time interval during the Bob's flight process, the transmit beamforming vectors for useful messages and AN projection are given by the rule of maximizing signal-to-leakage-and-noise ratio (Max-SLNR) and maximizing AN-and-leakage-to-noise ratio (ANLNR), and the optimal PA strategy is based on maximizing SR (Max-SR). More importantly, an alternating iterative structure (AIS) between beamforming and PA is proposed to further improve the SR performance. Simulation results show that the proposed AIS converges rapidly, and can achieve substantial SR gains over Max-SLNR plus Max-ANLNR with fixed PA such as PA factor $\beta=$ 0.5, and 0.9. In particular, in the case of small-scale antenna array, the SR performance gain achieved by the proposed AIS is more attractive. Additionally, as the number of antennas tends to be large-scale, the average SR performance of the proposed AIS approaches an SR ceil.
\end{abstract}


\IEEEpeerreviewmaketitle

\section{Introduction}
In recent years, unmanned aerial vehicle (UAV) has attracted a lot of interests in wireless communication due to its high mobility, flexible deployment and low cost\cite{Cheng2018UAV,Chen2018A,Zeng2017Energy,JointWu2018}. Because of the broadcast characteristic of wireless signals, secure wireless transmission in UAV networks is a very challenging problem\cite{Zhao2018Caching}. In the last decade, physical-layer security (PLS) has been heavily and widely investigated in wireless networks \cite{Wyner1975Wire,Wang2012Distributed,Guo2017Exploiting,Liu2017Secure,Ma2017Interference,Shu2017Artificial,Hu2016Artificial,Shu2017Secur e}. In \cite{Wyner1975Wire}, from information-theoretical viewpoint, Wyner first proved and claimed that secrecy capacity may be achievable if the eavesdropping channel is weaker than the desired one. Recently, there are some studies of focusing on the security of UAV systems. In \cite{Liu2017Secure}, the authors combined the transmission outage probability and secrecy outage probability as a performance metric to optimize the power allocation (PA) strategy. In addition, the authors in \cite{Wang2017Improving} utilized the UAV as a mobile relay node to improve secure transmission, where an iterative algorithm based on difference-of-concave program was also developed to circumvent the non-convexity of secrecy rate (SR) maximization.

Since the channel from ground base station Alice to UAV Bob can be usually regarded as line-of-sight (LoS) link, directional modulation (DM) technology can be naturally applied to UAV networks\cite{Lee2018UAV,Khuwaja2018A}. As one of the key techniques in PLS, DM is recently attracted an ever-increasing attention from both academia and industry world\cite{Babakhani2008Transmitter,Chen2017A,Zou2015Relay}. To enhance the PLS in DM networks, the artificial noise (AN) is usually exploited to degrade the eavesdropping channel\cite{Negi2005Secret,Goel2008Guaranteeing,Wu2017Secure}. The authors in \cite{Ding2014} first introduced the null-space projection method to project the AN along the eavesdropping direction on the null-space of the steering vector. In the multicast scenario, the leakage-based method is adopted to design both the precoding vector and AN projection matrix\cite{Shu2017Artificial}. In order to maximize the SR, a general power iterative (GPI) scheme was proposed in \cite{Yu2018GPI} to optimize both the useful precoding vector and the AN projection matrix. In DM systems, to implement the DM synthesis, the directional angle should be measured in advance. This generates measurement errors\cite{Shu2017Low}. In the presence of angle measurement errors, the authors in \cite{Hu2016Robust,Shu2017Robust} proposed two robust DM synthesis beamforming schemes to exploit the statical property of these errors. Actually, the two method achieved a substantial SR improvement over existing non-robust ones. And in the absence of the statistical distribution of measurement error, a blind robust main-lobe-integration-based leakage beamformer was proposed to achieve an improved security in DM systems\cite{Shu2018Robust}. In \cite{Hu2016Artificial}, a random frequency diverse (RFD) array was combined with directional modulation with the aid of AN to achieve a secure precise wireless transmission (SPWT). To reduce the circuit cost and complexity of SPWT receiver, the authors in \cite{Shu2017Secure}  replaced RFD with random subcarrier selection to achieve a SPWT.

Given the fixed beamforming vector and AN projection matrix, PA between confidential messages and AN will have a direct impact on SR performance as shown in the following literature. In \cite{Yang2015Artificial}, the authors first derived the secrecy outage probabilities (SOPs) for the on-off transmission scheme and the adaptive transmission scheme. Using the SOPs, the optimal PA was presented to maximize the SR. Given matched-filter precoder and the NSP of AN, a PA strategy of maximizing secrecy rate (Max-SR) was proposed and shown to make a significant SR gain in the case of small-scale antenna array and low signal-to-noise ratio (SNR) region in \cite{Wan2018Power}.

However, in the above literatures, the beamformimg scheme and PA are independently optimized. How to establish the relationship between them? In other words, how to implement the information propagation between them? In this paper, we propose an alternative iterative structure (AIS) between PA and beamforming with the aim of improving the average SR. During the UAV's flight from source to destination, the secrecy sum-rate (SSR) maximum problem is converted into independent subproblems per position. The subproblem of maximizing SR is a joint optimization of the beamforming vector, AN projection vector, and PA factor. This joint optimization problem is very hard. To simplify this problem, the beamforming and AN projection vectors are constructed by the leakage criterion. Then, we maximize SR to get the PA factor. Next, using the designed PA factor, the beamforming and AN projection vectors is computed again. This process is repeated until the terminal condition is satisfied. This forms the AIS proposed by us. From simulations, it follows that the proposed AIS converges quickly and achieves a substantial SR performance gain over fixed beamforming scheme plus fixed PA strategies.

The remainder of the paper is organized as follows. In Section II, we present the system model and the problem formulation. In Section III, the beamforming vector and AN projection vector are given, the PA strategy of Max-SR is derived, and the AIS is proposed. Simulation and numerical results are shown in Section IV. Finally, we draw our conclusions in Section V.

\emph{Notations:} Throughout the paper, matrices, vectors, and scalars are denoted by letters of bold upper case, bold lower case, and lower case, respectively. Signs $(\cdot)^T$, $(\cdot)^H$, $\mid\cdot\mid$ and $\parallel\cdot\parallel$ represent transpose, conjugate transpose, modulus and norm, respectively. $\textbf{I}_N$ denotes the $N\times N$ identity matrix.

\section{System Model}
\begin{figure}
  \centering
  \includegraphics[width=0.5\textwidth]{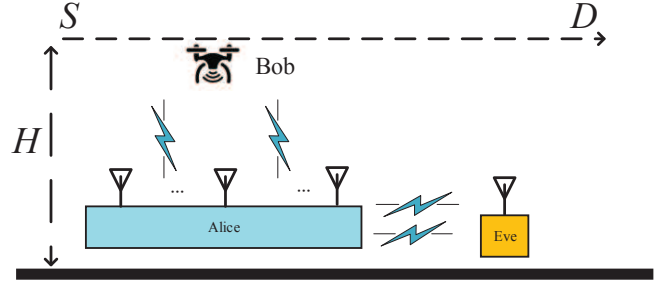}\\
  \caption{Schematic diagram of UAV system with directional modulation.}\label{Sys_Mod}
\end{figure}

In Fig.~\ref{Sys_Mod}, a UAV network is shown. Here, a legitimate UAV user (Bob) flies along an $L$-meters long direct link between S and D, while receives messages from the base station (Alice). On ground, Alice with $M$ transmit antennas performs as a base station, and sends confidential messages to Bob. Also there exists a potential illegal receiver Eve to wiretap confidential messages. We assume that Alice and Eve are both on the ground and are located at two fixed positions. Bob flies at an altitude of $H$ (m) above ground, with a constant moving speed $V$ (m/s), and the total flight time interval from $S$ to $D$ is $T$ (s).

As UAV Bob moves from S to D, we sample its position with equal time spacing $\Delta{t}$ and the total number of sampling points is $N=\frac{T}{\Delta{t}}$. The transmit signal at sampling point $n$ is expressed as
\begin{equation}\label{Tx signal s}
\mathbf{s}_n=\sqrt{\beta_n P_s}\mathbf{v}_{b,n}x_n+\sqrt{(1-\beta_n)P_s}\mathbf{v}_{AN,n}z_n
\end{equation}
where $P_s$ is the total transmit power, $n\in\mathcal{N}=\left\{1,2,\cdots,N\right\}$ denotes the $n$th sampling point, $\beta_n$ and  $(1-\beta_n)$ stand for the PA parameters for confidential messages and AN, respectively. $\mathbf{v}_{b,n}\in\mathbb{C}^{M\times1}$ denotes the transmit beamforming vector of the confidential message to the desired direction, and $\mathbf{v}_{AN,n}\in\mathbb{C}^{M\times1}$ is the beamforming vector of AN forwarded to the undesired direction, $\mathbf{v}^H_{b,n}\mathbf{v}_{b,n}=1$ and $\mathbf{v}^H_{AN,n}\mathbf{v}_{AN,n}=1$. In (\ref{Tx signal s}), $x_n$ is the confidential message satisfying $\mathbb{E}\left\{x_n^Hx_n\right\}=1$, and $z_n\sim\mathcal{C}\mathcal{N}(0,1)$ denotes the scalar AN being a complex Gaussian random variable with zero mean and unit variance.

The corresponding receive signal at Bob can be written as
\begin{align}\label{Rx_signal yb}
y_{b,n}
&=\mathbf{h}^{H}(\theta_{b,n})\mathbf{s}_n+w_{b,n}\nonumber\\
&=\sqrt{g_{ab}\beta_n P_s}\mathbf{h}^{H}(\theta_{b,n})\mathbf{v}_{b,n}x_n\nonumber\\
&+\sqrt{g_{ab}(1-\beta_n)P_s}\mathbf{h}^{H}(\theta_{b,n})\mathbf{v}_{AN,n}z_n+w_{b,n}
\end{align}
where $g_{ab}=\frac{\alpha}{d_{ab}^c}$ represents the path loss from Alice to Bob, $d_{ab}$ is the distance from Alice to Bob, $c$ is the path loss exponent and $\alpha$ is the path loss at reference distance $d_0$,  $\mathbf{h}(\theta_{b,n})\in\mathbb{C}^{M\times1}$ represents the steering vector from Alice to Bob, and $\theta_{b,n}$ is the direction angle from Alice to Bob at position $n$. The complex additive white Gaussian noise (AWGN) at Bob is denoted by $w_{b,n} \sim\mathcal{C}\mathcal{N}(0,\sigma_b^2)$.

In the same manner, the received signal at Eve is given by
\begin{align}\label{Rx_signal ye}
y_{e,n}
&=\mathbf{h}^{H}(\theta_{e,n})\mathbf{s}_n+w_{e,n}\nonumber\\
&=\sqrt{g_{ae}\beta_n P_s}\mathbf{h}^{H}(\theta_{e,n})\mathbf{v}_{b,n}x_n\nonumber\\
&+\sqrt{g_{ae}(1-\beta_n)P_s}\mathbf{h}^{H}(\theta_{e,n})\mathbf{v}_{AN,n}z_n+w_{e,n}
\end{align}
where $g_{ae}=\frac{\alpha}{d_{ae}^c}$, $g_{ae}$ denotes the path loss from Alice to Eve, $\mathbf{h}(\theta_{e,n})\in\mathbb{C}^{M\times1}$ denotes the steering vector from Alice to Eve, and $\theta_{e,n}$ means the direction angle from Alice to Eve at position $n$. The complex AWGN at Eve is $w_{e,n}\sim\mathcal{C}\mathcal{N}(0,\sigma_e^2)$. %For simplicity, we assume that $\sigma_b^2=\sigma_e^2=\sigma^2$.

It should be mentioned that the steering vectors in (\ref{Rx_signal yb}) and (\ref{Rx_signal ye}) have the following form
\begin{equation}\label{h_theta_b}
\mathbf{h}(\theta)=\left[e^{j2\pi\Psi_{\theta}(1)}, \cdots, e^{j2\pi\Psi_{\theta}(m)}, \cdots, e^{j2\pi\Psi_{\theta}(M)}\right]^T
\end{equation}
and the phase function $\Psi_{\theta}(m)$ is defined by
\begin{equation}\label{var_phi}
\Psi_{\theta}(m)\triangleq-\frac{(m-(M+1)/2)d\cos\theta}{\lambda}, m=1,2,\cdots, M
\end{equation}
where $m$ is the index of antenna, $\theta$ is the direction angle, $d$ represents the distance spacing between two adjacent antennas, and $\lambda$ is the carrier wavelength of transmit signal.

As per (\ref{Rx_signal yb}) and (\ref{Rx_signal ye}), the achievable rates from Alice to Bob and to Eve at sampling position $n$ can be expressed as
\begin{align}\label{Rb}
&R_{b,n}=\log_2\nonumber\\
&\left(1+\frac{g_{ab}\beta_n P_s
\mathbf{h}^{H}(\theta_{b,n})\mathbf{v}_{b,n}\mathbf{v}^H_{b,n}\mathbf{h}(\theta_{b,n})}{g_{ab}(1-\beta_n) P_s \mathbf{h}^{H}(\theta_{b,n})\mathbf{v}_{AN,n}\mathbf{v}^H_{AN,n}\mathbf{h}^{H}(\theta_{b,n})+\sigma^2_b}\right)\nonumber\\
\end{align}
and
\begin{align}\label{Re}
&R_{e,n}=\log_2\nonumber\\
&\left(1+\frac{g_{ae}\beta_n P_s
\mathbf{h}^{H}(\theta_{e,n})\mathbf{v}_{b,n}\mathbf{v}^H_{b,n}\mathbf{h}(\theta_{e,n})}{g_{ae}(1-\beta_n) P_s \mathbf{h}^{H}(\theta_{e,n})\mathbf{v}_{AN,n}\mathbf{v}^H_{AN,n}\mathbf{h}(\theta_{e,n})+\sigma^2_e}\right)\nonumber\\
\end{align}
respectively. Accordingly, the total achievable SSR $R_s$ of covering the flight from $S$ to $D$ can be written as
\begin{equation}\label{Rs}
R_s=\max\left\{0,\sum_{n=1}^NR_{b,n}-\sum_{n=1}^NR_{e,n}\right\}
\end{equation}

To maximize the above SSR $R_s$, we need to optimize the suitable beamforming vectors $\mathbf{v}_{b,n}$ and $\mathbf{v}_{AN,n}$, and PA factor $\beta_n$. Due to the independence of variables at each position $n$, the total achievable SSR maximization problem can be equivalently decomposed into $N$ subproblems of maximizing the SR at each sample point, that is
\begin{align}\label{Rsn}
\max_{\{\mathbf{v}_{b,n}, \mathbf{v}_{AN,n},\beta_n\}_{n=1}^N}& R_s=\sum_{n=1}^N \max_{\mathbf{v}_{b,n}, \mathbf{v}_{AN,n},\beta_n}R_{s,n}\\ \nonumber
&=\sum_{n=1}^N\max_{\mathbf{v}_{b,n}, \mathbf{v}_{AN,n},\beta_n}\max\left\{0, R_{b,n}-R_{e,n}\right\}
\end{align}

%and by solving the optimization problem of Max-SR to get the optimal power allocation variable $\beta^*_n$ at each sampling points.

Within the above right side summation, the $n$ term can be casted as the following complex joint optimization problem
\begin{align}\label{P1}
\mathrm{(P1):}&\max_{\mathbf{v}_{b,n}, \mathbf{v}_{AN,n}, \beta_n}~~~~R_{b,n}- R_{e,n}\nonumber\\
&~~\text{s. t.}~~0\leqslant\beta_n\leqslant1\nonumber\\
&~~~~~~~~\mathbf{v}^H_{b,n}\mathbf{v}_{b,n}=1\nonumber\\
&~~~~~~~~\mathbf{v}^H_{AN,n}\mathbf{v}_{AN,n}=1.
\end{align}

Obviously, it is very hard to solve the above joint optimization problem. Below, we first design the $\mathbf{v}_{b,n}$ and $\mathbf{v}_{AN,n}$ in advance by making use of rule of leakage. Provided that $\mathbf{v}_{b,n}$ and $\mathbf{v}_{AN,n}$ are given, the optimal PA strategy of Max-SR is addressed by the a simplified version of the above optimization problem. Due to the fixed values of $\mathbf{v}_{b,n}$ and $\mathbf{v}_{AN,n}$, the above joint optimization reduces to a single-variable PA problem as follows
\begin{align}\label{PA1}
\mathrm{(PA1):}&\max_{\beta_n}~~~~R_{s,n}(\beta_n)\nonumber\\
&~~\text{s. t.}~~0\leqslant\beta_n\leqslant1.
\end{align}
with $R_{s,n}(\beta_n)=\max\{0,R_{b,n}-R_{e,n}\}$.

\section{Proposed AIS scheme}
In this section, we first compute $\mathbf{v}_{b,n}$ and $\mathbf{v}_{AN,n}$ by using the concept of leakage. Following this, the optimal PA strategy of Max-SR will be solved by setting the first-order derivative to zero. Finally, an alternating iterative structure is established between $\{\mathbf{v}_{b,n}, \mathbf{v}_{AN,n}\}$ and $\beta_n$ to further improve the SR performance. Provided that the value of $\beta_n$ is available, the beamforming vectors $\mathbf{v}_{b,n}$ and $\mathbf{v}_{AN,n}$ are independently designed by the basic concept of leakage in \cite{Sadek2007A,Feng2011Ane} due to their independent property that the AN leakage from Eve to Bob and the confidential message leakage from Bob to Eve.

\subsection{Design $\mathbf{v}_{b,n}$ for fixed $\beta_n$}
From the aspect of Bob, we design the beamforming vector $\mathbf{v}_{b,n}$ of the useful information-carried signal by minimizing its leakage to Eve. It can be expressed by the following optimization problem
\begin{align}\label{P2}
\mathrm{(P2):}&\max_{\mathbf{v}_{b,n}}~~~~\mathrm{SLNR}(\mathbf{v}_{b,n}, \text{fixed}~\beta_n)\nonumber\\
&~~\text{s. t.}~~\mathbf{v}^H_{b,n}\mathbf{v}_{b,n}=1
\end{align}
where
\begin{align}
&\mathrm{SLNR}(\mathbf{v}_{b,n}, \text{fixed}~\beta_n)=\\ \nonumber
&\frac{\beta_n P_s\mathbf{v}^H_{b,n}\mathbf{h}(\theta_{b,n})\mathbf{h}^{H}(\theta_{b,n})\mathbf{v}_{b,n}}{\mathbf{v}^H_{b,n}(\beta_n P_s\mathbf{h}(\theta_{e,n})\mathbf{h}^{H}(\theta_{e,n})+\sigma_b^2\mathbf{I}_{N})\mathbf{v}_{b,n}}
\end{align}

Using the generalized Rayleigh-Ritz ratio theorem, the optimal $\mathbf{v}_{b,n}$ for maximizing the SLNR can be obtained from the eigenvector corresponding to the largest eigen-value of the matrix
\begin{align}\label{slnr_v_b}
\left[\mathbf{h}(\theta_{e,n})\mathbf{h}^{H}(\theta_{e,n})\beta_n P_s+\sigma_b^2
\mathbf{I}_{N}\right]^{-1}\mathbf{h}(\theta_{b,n})\mathbf{h}^{H}(\theta_{b,n})
\end{align}

Since the rank of the above matrix is one, we can directly give the closed-form solution to (\ref{P2}) as
\begin{align}\label{slo_vb}
\mathbf{v}_{b,n}=\frac{[\mathbf{h}(\theta_{e,n})\mathbf{h}^{H}(\theta_{e,n})\beta_n P_s+\sigma^2_b
\mathbf{I}_{N}]^{-1}\mathbf{h}(\theta_{b,n})}{\|[\mathbf{h}(\theta_{e,n})\mathbf{h}^{H}(\theta_{e,n})\beta_n P_s+\sigma^2_b
\mathbf{I}_{N}]^{-1}\mathbf{h}(\theta_{b,n})\|_2}
\end{align}

\subsection{Design $\mathbf{v}_{AN,n}$ for fixed $\beta_n$}
Similar to $\mathbf{v}_{b,n}$, from the aspect of Eve, we design the beamforming vector $\mathbf{v}_{AN,n}$ of AN by minimizing its leakage to Bob, called maximizing AN-and-leakage-to-noise ratio (ANLNR), which is formed as
\begin{align}\label{P}
\mathrm{(P3):}&\max_{\mathbf{v}_{AN,n}}~~~~\mathrm{ANLNR}(\mathbf{v}_{AN,n}, \text{fixed}~\beta_n)\nonumber\\
&~~\text{s. t.}~~\mathbf{v}^H_{AN,n}\mathbf{v}_{AN,n}=1
\end{align}
where
\begin{align}
&\mathrm{ANLNR}(\mathbf{v}_{AN,n}, \text{fixed}~\beta_n)=\nonumber\\
&\frac{(1-\beta_n) P_s\mathbf{v}^H_{AN,n}\mathbf{h}(\theta_{e,n})\mathbf{h}^{H}(\theta_{e,n})\mathbf{v}_{AN,n}}
{\mathbf{v}^H_{AN,n}((1-\beta_n) P_s\mathbf{h}(\theta_{b,n})\mathbf{h}^{H}(\theta_{b,n})+\sigma_e^2\mathbf{I}_{N})\mathbf{v}_{AN,n}}
\end{align}

In accordance with the generalized Rayleigh-Ritz ratio theorem, the optimal $\mathbf{v}_{AN,n}$ for maximizing the ANLNR is also obtained from the eigenvector corresponding to the largest eigen-value of the matrix
\begin{align}\label{slnr_v_an}
\left[\mathbf{h}(\theta_{b,n})\mathbf{h}^{H}(\theta_{b,n})(1-\beta_n)P_s+\sigma^2_e \mathbf{I}_{N}\right]^{-1}\mathbf{h}(\theta_{e,n})\mathbf{h}^{H}(\theta_{e,n})
\end{align}
which further yields
\begin{align}\label{slo_van}
\mathbf{v}_{AN,n}=\frac{[\mathbf{h}(\theta_{b,n})\mathbf{h}^{H}(\theta_{b,n})(1-\beta_n)P_s+\sigma_e^2 \mathbf{I}_{N}]^{-1}\mathbf{h}(\theta_{e,n})}{\|[\mathbf{h}(\theta_{b,n})\mathbf{h}^{H}(\theta_{b,n})(1-\beta_n)P_s+\sigma_e^2 \mathbf{I}_{N}]^{-1}\mathbf{h}(\theta_{e,n})\|_2}.
\end{align}

\subsection{Optimize $\beta_n$ by the Max-SR rule for fixed $\mathbf{v}_{b,n}$ and $\mathbf{v}_{AN,n}$}
In Subsections A and B, both $\mathbf{v}_{b,n}$ and $\mathbf{v}_{AN,n}$ are obtained by fixing the PA factor $\beta_n$. Now, with the known values of $\mathbf{v}_{b,n}$ and $\mathbf{v}_{AN,n}$, by introducing the auxiliary variable $t$, the optimization problem (\ref{PA1}) will be casted as

\begin{align}\label{PA2}
\mathrm{(PA2):}&\max_{\beta_n}~~~~t\nonumber\\
&~~\text{s. t.}~~t=\max\{0,f(\beta_n)\}\nonumber\\
&~~~~~~~~\beta_n\in [0,~1].
\end{align}

where
\begin{align}\label{Rs_beta_1}
&f(\beta_n)=R_{b,n}-R_{e,n}=\log_2\nonumber\\
&\left(1+\frac{g_{ab}\beta_n P_s
\mathbf{h}^{H}(\theta_{b,n})\mathbf{v}_{b,n}\mathbf{v}^H_{b,n}\mathbf{h}(\theta_{b,n})}{g_{ab}(1-\beta_n) P_s \mathbf{h}^{H}(\theta_{b,n})\mathbf{v}_{AN,n}\mathbf{v}^H_{AN,n}\mathbf{h}^{H}(\theta_{b,n})+\sigma^2_b}\right)\nonumber\\
&-\log_2\nonumber\\
&\left(1+\frac{g_{ae}\beta_n P_s
\mathbf{h}^{H}(\theta_{e,n})\mathbf{v}_{b,n}\mathbf{v}^H_{b,n}\mathbf{h}(\theta_{e,n})}{g_{ae}(1-\beta_n) P_s \mathbf{h}^{H}(\theta_{e,n})\mathbf{v}_{AN,n}\mathbf{v}^H_{AN,n}\mathbf{h}(\theta_{e,n})+\sigma^2_e}\right)\nonumber\\
&=\log_2\frac{A\beta^2_n+B\beta_n+C}{D\beta^2_n+E\beta_n+F}\nonumber\\
&=\log_2\phi(\beta_n)
\end{align}
where
\begin{align}
A&=g_{ab}g_{ae}P_s^2 \|\mathbf{h}^{H}(\theta_{e,n})\mathbf{v}_{AN,n}\|_2^2\nonumber\\
&\times\left(\|\mathbf{h}^{H}(\theta_{b,n})\mathbf{v}_{AN,n}\|_2^2
-\|\mathbf{h}^{H}(\theta_{b,n})\mathbf{v}_{b,n}\|_2^2\right),\\
B&=(g_{ae}P_s\|\mathbf{h}^{H}(\theta_{e,n})\mathbf{v}_{AN,n}\|_2^2+\sigma_e^2)\nonumber\\
&\times g_{ab}P_s\left(\|\mathbf{h}^{H}(\theta_{b,n})\mathbf{v}_{b,n}\|_2^2-\|\mathbf{h}^{H}(\theta_{b,n})\mathbf{v}_{AN,n}\|_2^2\right)-g_{ae}P_s\nonumber\\
&\|\mathbf{h}^{H}(\theta_{e,n})\mathbf{v}_{AN,n}\|_2^2
\left(g_{ab}P_s\|\mathbf{h}^{H}(\theta_{b,n})\mathbf{v}_{AN,n}\|_2^2+\sigma_b^2\right),\\
C&=\left(g_{ab}P_s\|\mathbf{h}^{H}(\theta_{b,n})\mathbf{v}_{AN,n}\|_2^2+\sigma_b^2\right)\nonumber\\
&\times\left(g_{ae}P_s\|\mathbf{h}^{H}(\theta_{e,n})\mathbf{v}_{AN,n}\|_2^2+\sigma_e^2\right),\\
D&=g_{ab}g_{ae}P_s^2 \|\mathbf{h}^{H}(\theta_{b,n})\mathbf{v}_{AN,n}\|_2^2\nonumber\\
&\times\left(\|\mathbf{h}^{H}(\theta_{e,n})\mathbf{v}_{AN,n}\|_2^2
-\|\mathbf{h}^{H}(\theta_{e,n})\mathbf{v}_{b,n}\|_2^2\right),\\
E&=(g_{ab}P_s\|\mathbf{h}^{H}(\theta_{b,n})\mathbf{v}_{AN,n}\|_2^2+\sigma_b^2)\nonumber\\
&\times g_{ae}P_s\left(\|\mathbf{h}^{H}(\theta_{e,n})\mathbf{v}_{b,n}\|_2^2-\|\mathbf{h}^{H}(\theta_{e,n})\mathbf{v}_{AN,n}\|_2^2\right)-g_{ab}P_s\nonumber\\
&\|\mathbf{h}^{H}(\theta_{b,n})\mathbf{v}_{AN,n}\|_2^2
\left(g_{ae}P_s\|\mathbf{h}^{H}(\theta_{e,n})\mathbf{v}_{AN,n}\|_2^2+\sigma_e^2\right)
\end{align}
with $F=C$.

Optimization problem (\ref{PA2}) is a non-convex program due to the first constraint, and it can be solved by one-dimensional exhaustive search method (ESM). To lower the complexity of the one-dimensional ESM, we propose a closed-form solution to optimization problem problem (\ref{PA2}). Considering $f(\beta_n)=0$ at $\beta_n=0$, the first constraint can be rewritten as $t=\max\{f(0),f(\beta_n)\}$, which forms
\begin{align}\label{PA3}
\mathrm{(PA3):}&\max_{\beta_n}~~~~f(\beta_n)\nonumber\\
&~~\text{s. t.}~~\beta_n\in [0,~1].
\end{align}

Observing the definition of function $f(\beta_n)$ shown in (\ref{Rs_beta_1}), it is very clear that $f(\beta_n)$ is a continuous and differentiable function of variable $\beta_n$ in the closed interval $[0,~1]$. Thus, the optimal $\beta_n^*$ must locate at some endpoints or some stationary points. In what follows, we solve this problem in two steps: firstly, find the stationary points by vanishing the first-order derivative of $R_{s,n}$; secondly, select the optimal value of $\beta_n$ by comparing the values of $R_{s,n}$ among the set of candidate points to the critical number.

The critical points of $R_{s,n}$ can be solved by
\begin{equation}\label{Rs_beta_deri}
\frac{\partial R_{s,n}}{\partial\beta_n}=0,
\end{equation}
which is further reduced to
\begin{align}
&\frac{\partial \phi(\beta_n)}{\partial\beta_n}=\nonumber\\
&\frac{(AE-BD)\beta^2_n+2C(A-D)\beta_n+C(B-E)}{(D\beta^2_n+E\beta_n+C)^2}=0,
\end{align}
which yields
\begin{align}\label{beta_1}
\beta_{n,1}=\frac{-C(A-D)+\sqrt{\Delta}}{AE-BD},
\end{align}
and
\begin{align}\label{beta_2}
\beta_{n,2}=\frac{-C(A-D)-\sqrt{\Delta}}{AE-BD},
\end{align}
where
\begin{equation}
\Delta=C^2(A-D)^2-C(AE-BD)(B-E)
\end{equation}

In summary, considering $\beta\in[0,1]$, we have the candidate set for the critical number of function $f(\beta_n)$ as follows
\begin{align}
S_C=\left\{0,\beta_{n,1}, \beta_{n,2}, 1\right\}.
\end{align}

In what follows, we need to decide which one in set $S_C$ is the final solution to maximize the function $f(\beta_n)$. Obviously, $\beta_n=0$ means that $f(0)=0$. That is, the SR is equal to zero. Thus, the $\beta_n=0$ should be deleted from the above candidate set. We have a reduced candidate set as follows
\begin{align}\label{RCS}
\tilde{S}_C=\left\{\beta_{n,1}, \beta_{n,2}, 1\right\}.
\end{align}

 Below, let us discuss these two stationary points $\beta_{n,1}$, and $\beta_{n,2}$ under what condition of $\Delta$, i.e., the sign of value of $\Delta$. The first special case is $\Delta<0$, then the two real roots $\beta_{n,1}$ and $\beta_{n,2}$ will not exist. On this basis, we should check the remaining three cases as follows.\\
\textbf{Case 1.} $AE-BD>0$, the rational function $\phi(\beta_n)$ is a monotonously increasing function. It will achieve the maximum value at $\beta_n=1$. \\
\textbf{Case 2.} $AE-BD=0$, the stationary point is $\beta_{n,3}=\frac{E-B}{2(A-D)}$. We need to judge whether $\beta_{n,3}\in(0,1)$. While $\beta_{n,3}\in(0,1)$, we obtain the PA parameter $\beta_n$ by comparing the value of $\phi(\beta_{n,3})$ and $\phi(1)$. Otherwise, the optimal PA factor is selected as $\beta_n=1$.\\
\textbf{Case 3.} $AE-BD<0$, the rational  function $\phi(\beta_n)$ is a monotonously decreasing function. Therefore, the PA parameter is $\beta_n=0$. This result leads to a contradiction with the reduced candidate set in (\ref{RCS}).

As for $\Delta\geqslant 0$, we need to judge whether the two candidates meeting the conditions that the PA parameter lies in the interval of (0, 1). Then, compare the values of $\phi(\beta_n)$ at the endpoints and corresponding stationary points to get the optimal PA parameter. There are four different cases.\\
\textbf{Case 1.} If $\beta_{n,1}\in(0,1)$, $\beta_{n,2}\in(0,1)$, then compare the values of $\phi(\beta_{n,1})$, $\phi(\beta_{n,2})$ and $\phi(1)$.\\
\textbf{Case 2.} If $\beta_{n,1}\in(0,1)$, $\beta_{n,2}\notin(0,1)$, then compare the values of $\phi(\beta_{n,1})$ and $\phi(1)$.\\
\textbf{Case 3.} If $\beta_{n,1}\notin(0,1)$, $\beta_{n,2}\in(0,1)$, then compare the values of $\phi(\beta_{n,2})$ and $\phi(1)$.\\
\textbf{Case 4.} If $\beta_{n,1}\notin(0,1)$, $\beta_{n,2}\notin(0,1)$, then the value of $\phi(1)$ will be optimal.

After making the above comparison, we can get the optimal PA parameter $\beta_n$ of Max-SR given the values of $\mathbf{v}_{b,n}$ and $\mathbf{v}_{AN,n}$.

\subsection{Proposed AIS}

\begin{figure}
  \centering
  \includegraphics[width=0.45\textwidth]{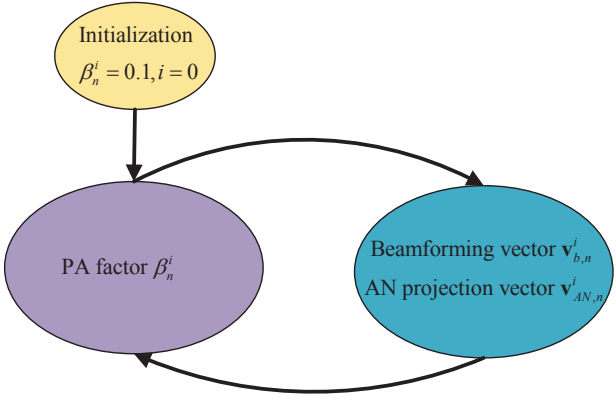}\\
  \caption{Proposed AIS}\label{fig3}
\end{figure}

\begin{figure}
  \centering
  \includegraphics[width=0.45\textwidth]{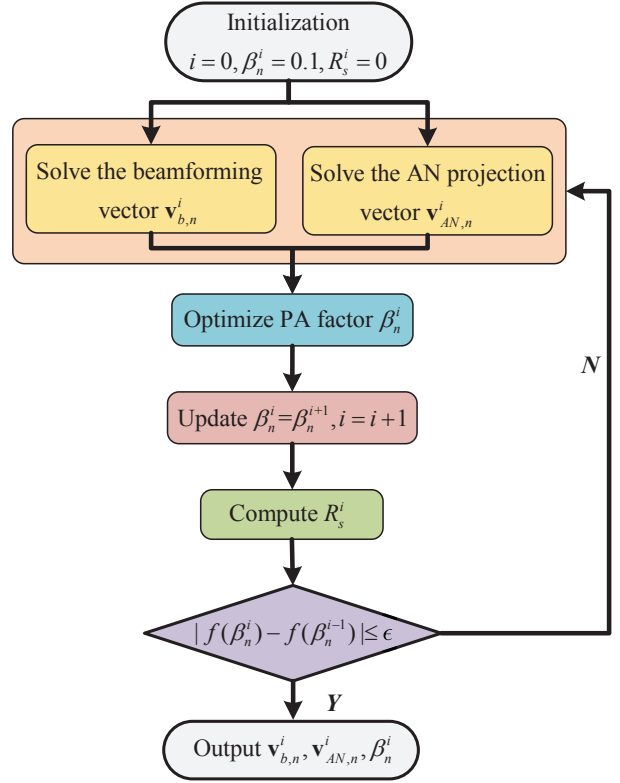}\\
  \caption{Flow chart of the proposed AIS}\label{fig22}
\end{figure}

To further enhance the SR performance in our system, an AIS sketched in Fig. \ref{fig3} is established among $\mathbf{v}^i_{b,n}$, $\mathbf{v}^i_{AN,n}$, and $\beta^i_n$ with an initial value of $\beta^0_n$, where superscript $i$ denotes the $i$th iteration of position $n$. Then, the PA parameter $\beta^1_n$ is decided  from several candidates via the discussion of different cases according to the process in Subsection C. Subsequently, using the new value of $\beta^1_n$, we compute the values of $\mathbf{v}^1_{b,n}$ and $\mathbf{v}^1_{AN,n}$ based on (\ref{slo_vb}) and (\ref{slo_van}). This process will be repeated until $f(\beta^{i}_{n})-f(\beta^{i-1}_n)$ is smaller than a predefined value. The detailed procedure is also indicated in Fig. \ref{fig22}.

To make clear, the iterative algorithm is summarized as Algorithm~\ref{algorithm 1}.
\begin{algorithm}
Initialization: $i=0, \beta^i_n=0.1, R_s^{i}=0$.
\begin{enumerate}
  \item For given $\beta^i_n$, compute the $\mathbf{v}^i_{b,n}$ and $\mathbf{v}^i_{AN,n}$ based on (\ref{slo_vb}) and (\ref{slo_van}),
  \item Solve (\ref{PA3}) and discuss aforementioned two scenarios to obtain $\beta^{i+1}_n$,
  \item Update $\beta^i_n=\beta^{i+1}_n, i=i+1$,
  \item Compute $R_s^{i}$,
  \item Until $|f(\beta^{i}_{n})-f(\beta^{i-1}_n)|\leq\epsilon$.
\end{enumerate}
\caption{Proposed alternating iterative algorithm}\label{algorithm 1}
\end{algorithm}

\section{Simulation and Discussion}
To evaluate the SR performance of the proposed AIS, simulation results and analysis are presented in the following. The parameters and specifications are used as follows: the spacing between two adjacent antennas is $d=\lambda/2$, the path loss exponent $c=2$, the distance between $S$ and $D$ is $L$=800m, the distance between Alice and Eve is 200m, the flying altitude of UAV Bob is $H$=20m, the UAV speed is $v$=8m/s, the sample interval $\Delta{t}$=1s, and the number of sampling points ${N}=\lfloor T/\Delta{t}\rfloor$.

\begin{figure}[!t]
  \centering
  \includegraphics[width=0.5\textwidth]{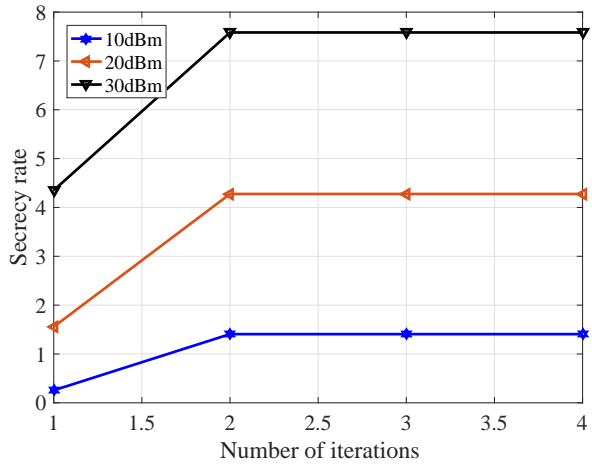}\\
  \caption{SR versus number of iterations for the proposed AIS.}\label{iterations_dBm}
\end{figure}

Fig.~\ref{iterations_dBm} demonstrates the convergence of the proposed AIS. From this figure, we can clearly observe that the proposed AIS can converge rapidly for three distinct transmit powers. It is very obvious that the proposed AIS can converges within one or two iterations. This convergence rate is attractive. More importantly, after convergence, the proposed AIS may achieve an excellent SR improvement before convergence.

\begin{figure}[!t]
  \centering
  \includegraphics[width=0.5\textwidth]{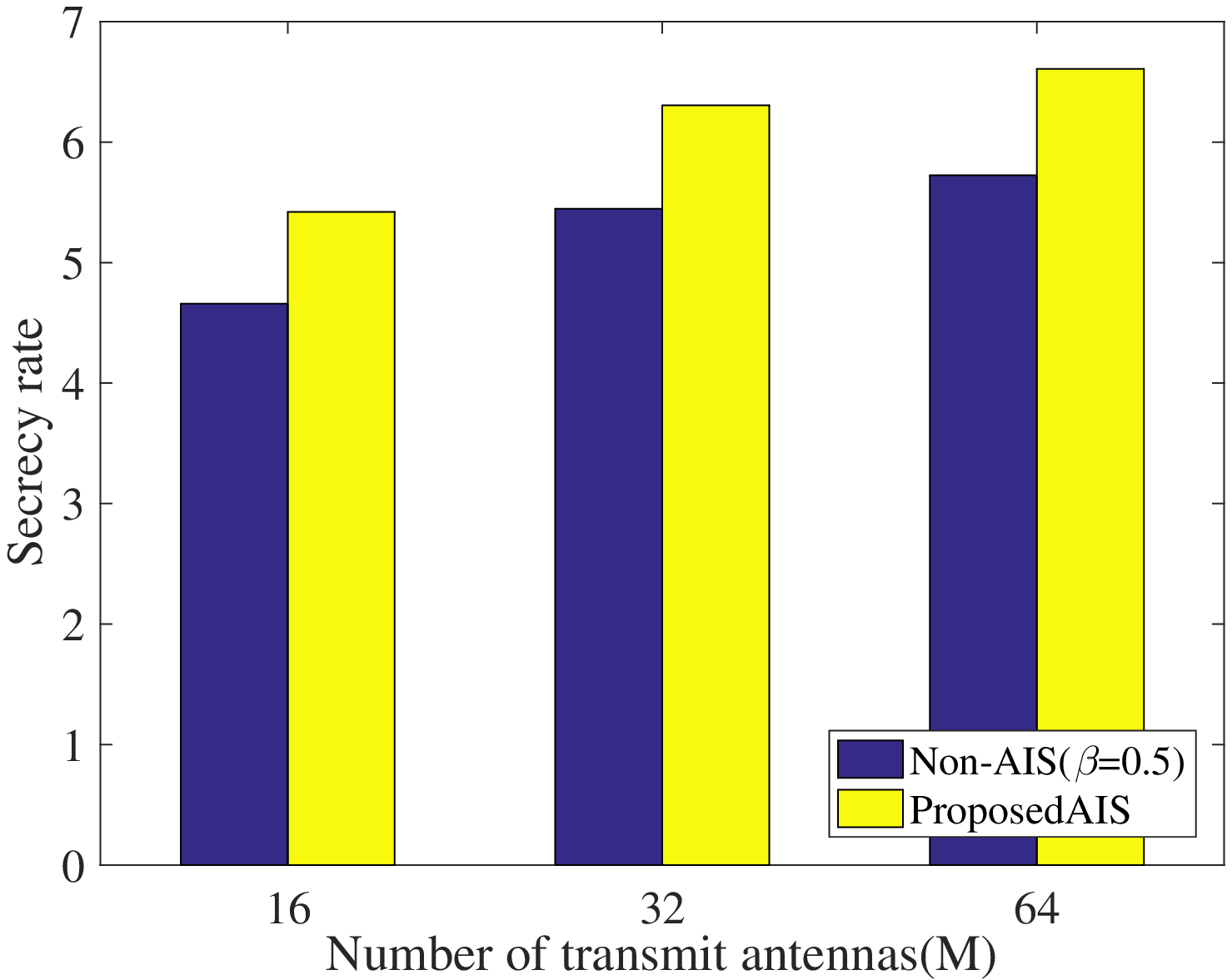}\\
  \caption{Histogram of SR versus number of antennas ($\beta=0.5$ and $P_s$=10dBm).}\label{beta_0.5_dbm_10}
\end{figure}

\begin{figure}[!t]
  \centering
  \includegraphics[width=0.5\textwidth]{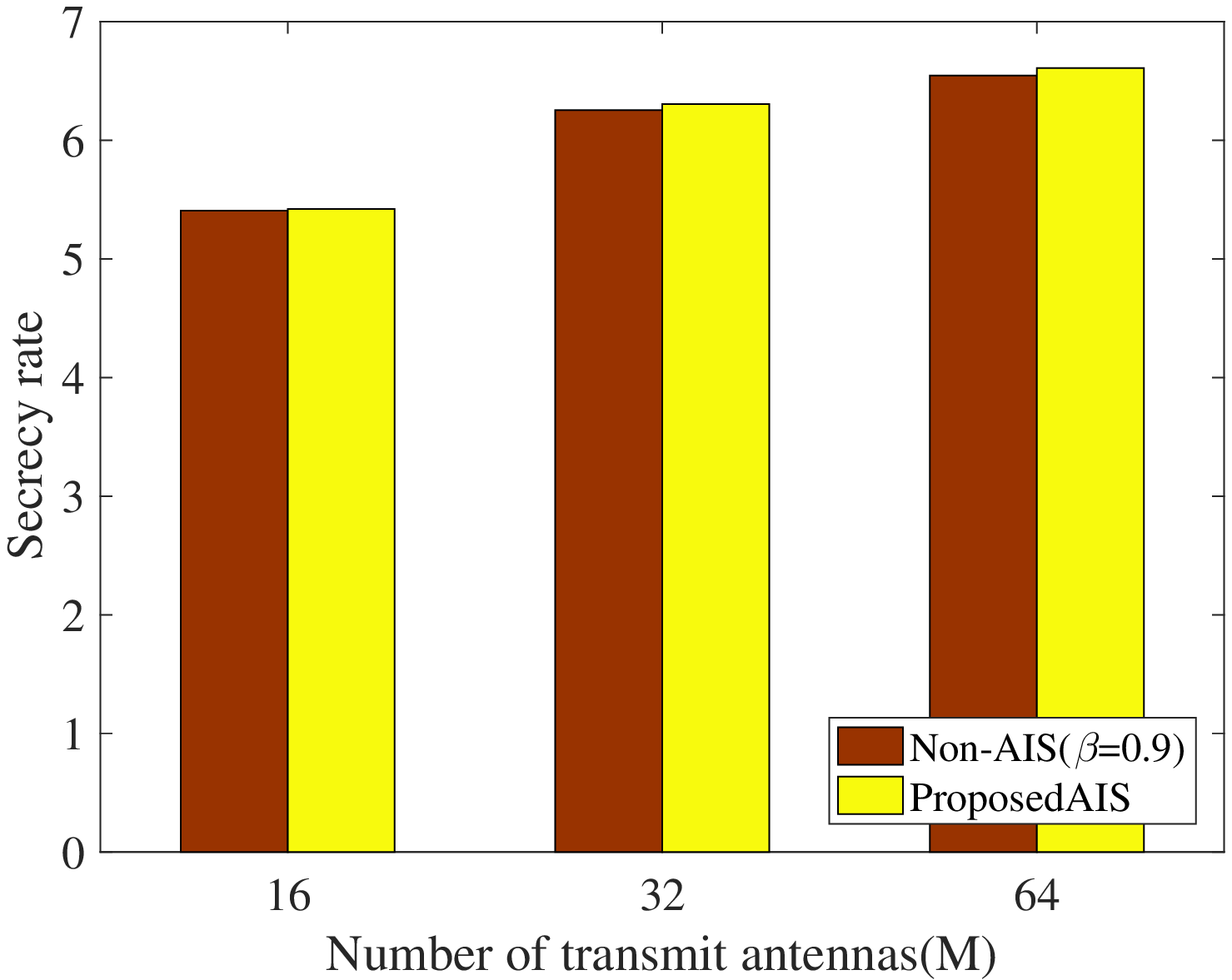}\\
  \caption{Histogram of SR versus number of antennas ($\beta=0.9$ and $P_s$=10dBm).}\label{beta_0.9_dbm_10}
\end{figure}

\begin{figure}[!t]
  \centering
  \includegraphics[width=0.5\textwidth]{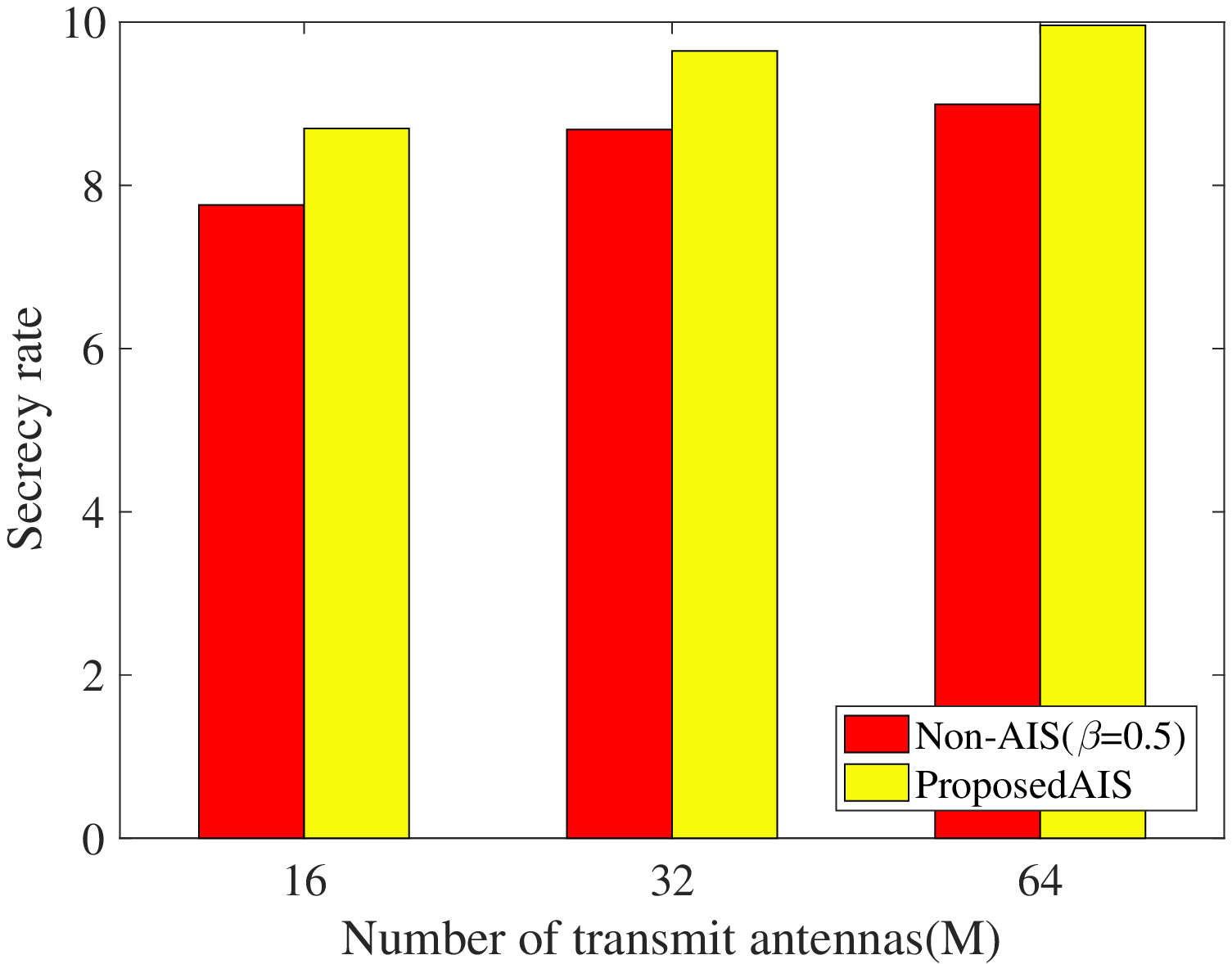}\\
  \caption{Histogram of SR versus number of antennas ($\beta=0.5$ and $P_s$=20dBm).}\label{beta_0.5_dbm_20}
\end{figure}

\begin{figure}[!t]
  \centering
  \includegraphics[width=0.5\textwidth]{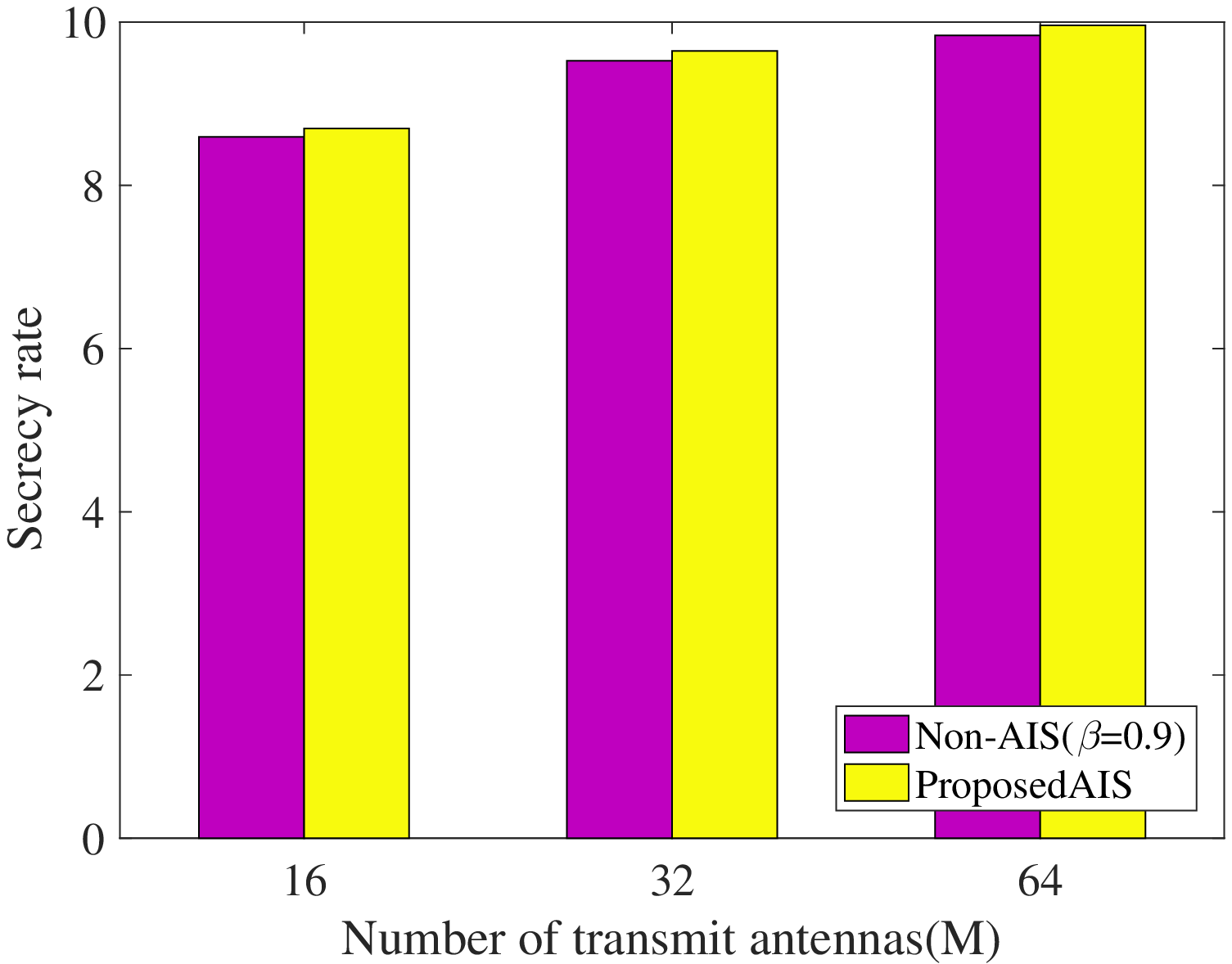}\\
  \caption{Histogram of SR versus number of antennas ($\beta=0.9$ and $P_s$=20dBm).}\label{beta_0.9_dbm_20}
\end{figure}

\begin{figure}[!t]
  \centering
  \includegraphics[width=0.5\textwidth]{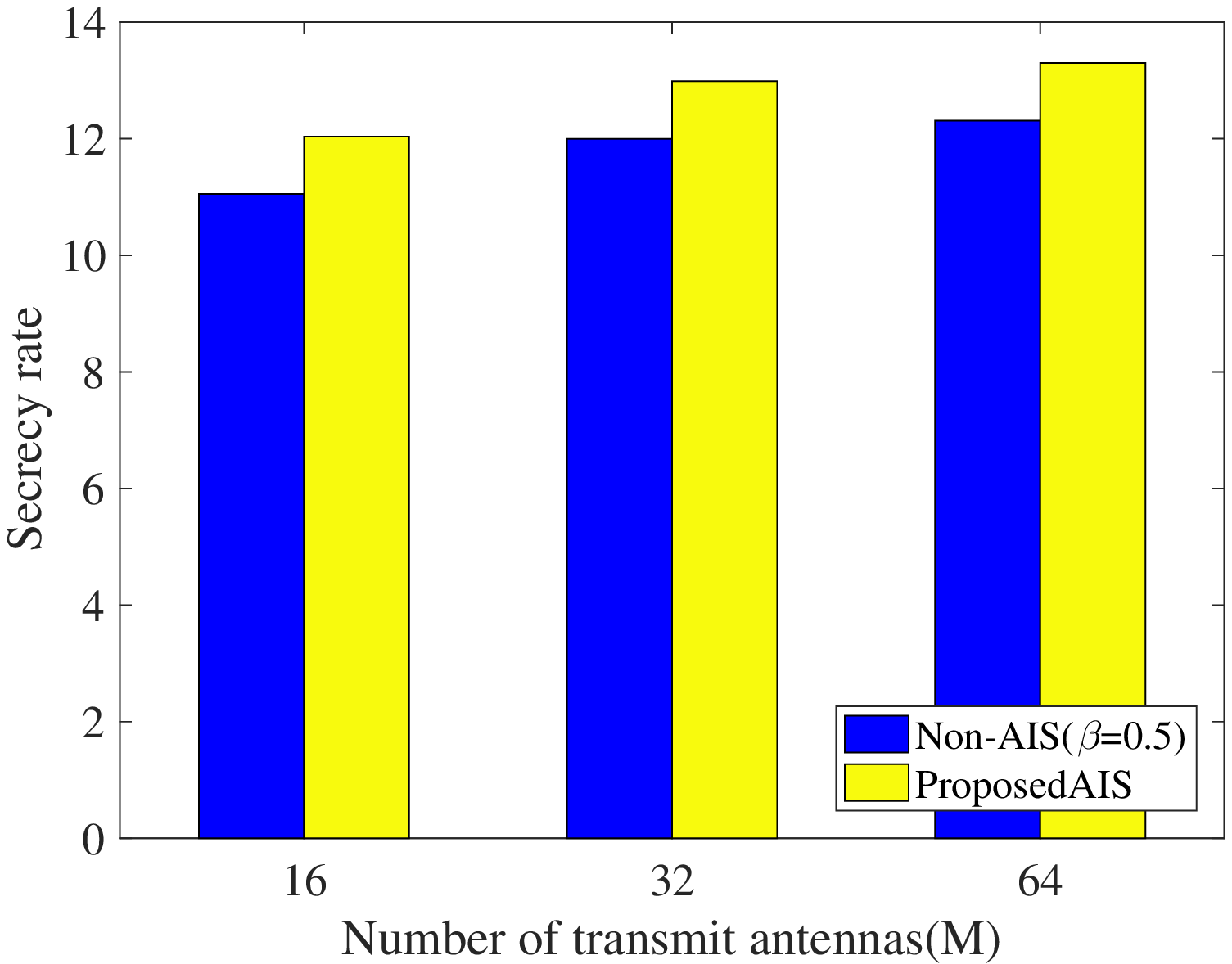}\\
  \caption{Histogram of SR versus number of antennas ($\beta=0.5$ and $P_s$=30dBm).}\label{beta_0.5_dbm_30}
\end{figure}

\begin{figure}[!t]
  \centering
  \includegraphics[width=0.5\textwidth]{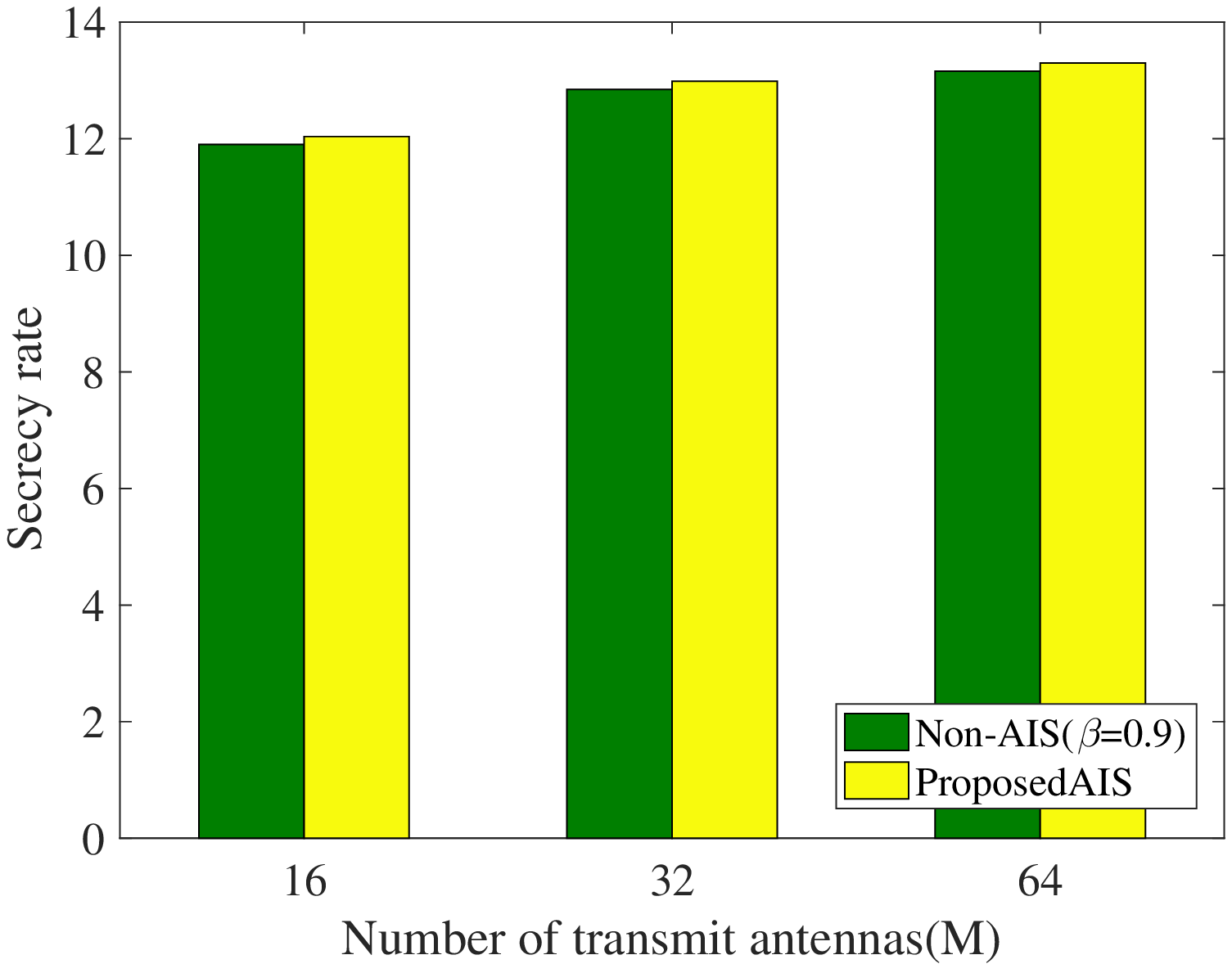}\\
  \caption{Histogram of SR versus number of antennas ($\beta=0.9$ and $P_s$=30dBm).}\label{beta_0.9_dbm_30}
\end{figure}

\begin{figure}[!t]
  \centering
  \includegraphics[width=0.5\textwidth]{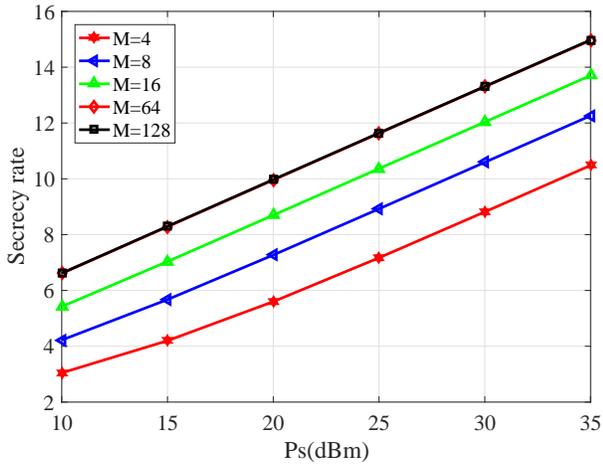}\\
  \caption{SR versus $P_s$ of the proposed AIS.}\label{SR_SNR}
\end{figure}

\begin{figure}[!t]
  \centering
  \includegraphics[width=0.5\textwidth]{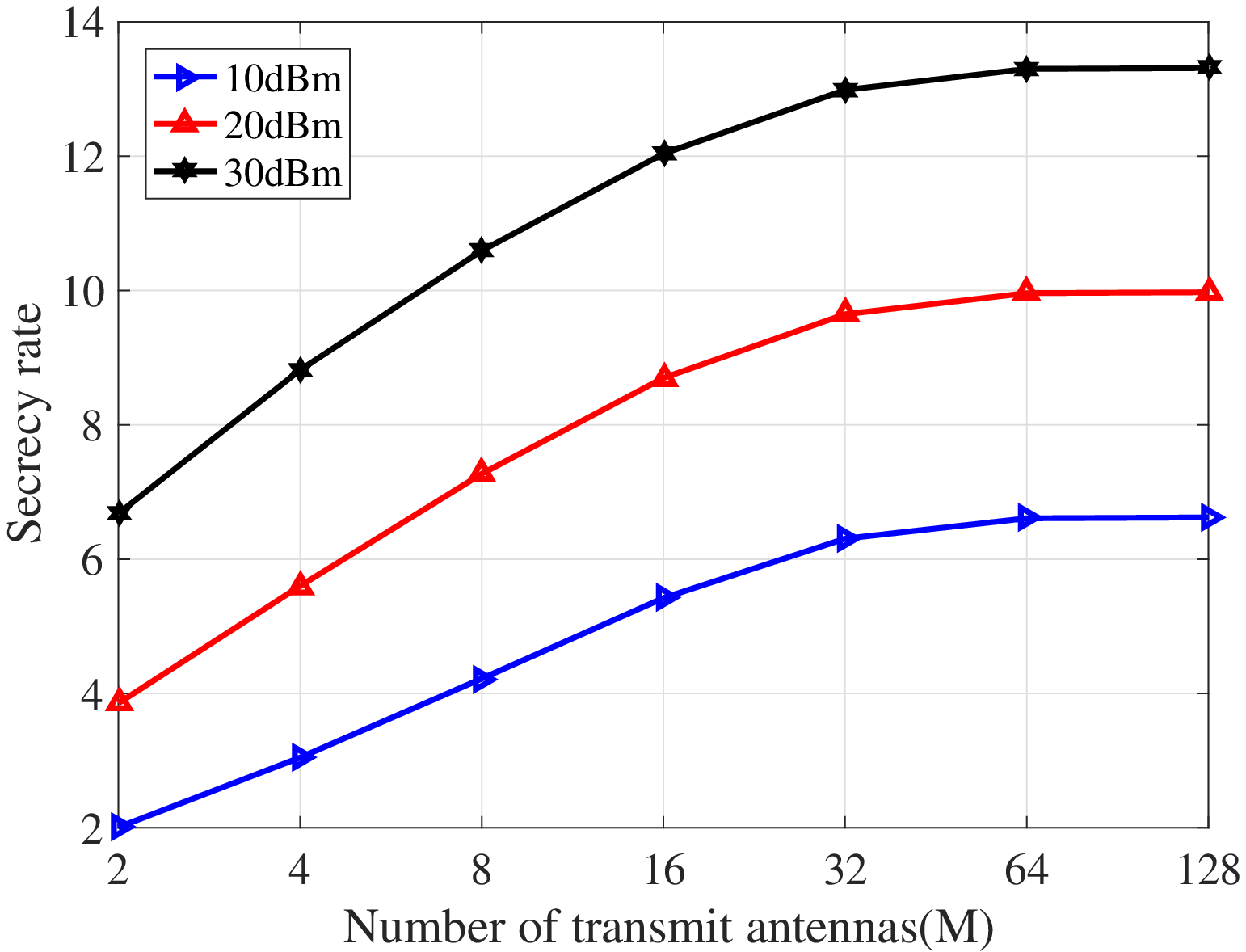}\\
  \caption{SR versus number of antennas of the proposed AIS.}\label{SR_SNR_dBm}
\end{figure}

Fig.~\ref{beta_0.5_dbm_10} and Fig.~\ref{beta_0.9_dbm_10} show the histograms of SR versus number of transmit antennas $M$ of the proposed AIS with optimal PA parameter $\beta^*$ compared with Max-SLNR plus Max-ANLNR with typical fixed $\beta=0.5$ and $\beta=0.9$ for $P_s$=10dBm, respectively. From the two figures, we can observe that the proposed AIS achieves a substantial and slight SR performance gains over Max-SLNR plus Max-ANLNR with fixed $\beta=0.5$, and $\beta=0.9$, respectively.

Next, in Fig.~\ref{beta_0.5_dbm_20}, and Fig.~\ref{beta_0.9_dbm_20}, we increase $P_s$ up to 20dBm. From the two figures, it can be clearly seen that the proposed AIS still shows an appreciated improvements over Max-SLNR plus Max-ANLNR with typical fixed $\beta=0.5$ and $\beta=0.9$, respectively.

Lastly, in Fig.~\ref{beta_0.5_dbm_30}, and Fig.~\ref{beta_0.9_dbm_30}, the transmit power $P_s$ is increased up to 30dBm. From the two figures, it can be clearly seen that the proposed AIS still shows an appreciated improvements over Max-SLNR plus Max-ANLNR with typical fixed $\beta=0.5$ and $\beta=0.9$, respectively.

Fig.~\ref{SR_SNR} plots the curves of SR versus $P_s$ for the proposed AIS with different numbers of transmit antennas. From this figure, we can see that the SR increases gradually as the transmit power $P_s$ for the fixed number of transmit antennas. Similarly, if we fix the transmit power, we find that increasing the number of transmit antennas will also improve the SR performance obviously. However, for a large number of antennas, the SR performance gain achieved by doubling the number of antennas becomes smaller.

Fig.~\ref{SR_SNR_dBm} illustrates the curves of SR versus number of transmit antennas of the proposed AIS. In Fig.~\ref{SR_SNR_dBm}, given  three different $P_s$ scenarios $P_S$=10dBm, 20dBm, and 30dBm, the SR performance approaches three different ceils as the number of transmit antennas tends to be large-scale.

\section{Conclusion}
In our work, we have investigated a UAV-enabled wireless system. An AIS is proposed to realize an iterative operation between beamforming and PA to further improve SR. Firstly, we established a complex joint optimization problem of maximizing SR. To make the complex joint optimization problem more simple, the Max-SLNR and Max-ANLNR criterion is adopted to construct the beamforming vector $\mathbf{v}_{b,n}$ and the AN projection vector $\mathbf{v}_{AN,n}$. Then, given the solved $\mathbf{v}_{b,n}$ and $\mathbf{v}_{AN,n}$, we turn to address the single-variable PA optimization problem of Max-SR. Actually, SR is regarded as a continuous and differentiable function of PA factor $\beta_n$ with $\beta_n$ being in a closed interval $[0,~1]$. By the analysis of the set of critical points, we can attain the optimal value of $\beta_n$. Finally, a low-complexity AIS between the beamforming and AN projection vectors, and the PA factor are proposed to further enhance the secrecy rate. Simulation results show that the proposed AIS can converge rapidly, and the optimal PA strategy can substantially improve the SR performance compared with some typical PA factors such as $\beta=0.5$, and $\beta=0.9$. Moreover, the SR of the proposed AIS grows gradually with the increasing of the $P_s$. Furthermore, our proposed AIS has achieved an appreciated SR performance gain in small-scale number of transmit antennas.

\ifCLASSOPTIONcaptionsoff
  \newpage
\fi

\bibliographystyle{IEEEtran}
\bibliography{IEEEfull,mycite}

\end{document}